\documentclass[trackchanges, twocolumn]{aastex701}
\usepackage{amsmath}
\usepackage{bm}
\usepackage{booktabs}
\usepackage{microtype}

\newcommand{\RX}{\mathcal{R}_{X}}
\newcommand{\QX}{\mathcal{Q}_{X}}
\newcommand{\MX}{\mathcal{M}_{X}}
\newcommand{\Crad}{\mathcal{C}_{\rm rad}}
\newcommand{\FIC}{\mathcal{F}_{\rm IC}}
\newcommand{\Katm}{\mathcal{K}_{\rm atm}}
\newcommand{\dd}{\mathrm{d}}
\newcommand{\e}{\mathrm{e}}

\begin{document}

\hypersetup{pdftitle={The Conditional CMB Visibility of High-Redshift X-Ray Cavities in Galaxy clusters},pdfauthor={Ramananda Santra, Samik Mitra}}

\title{The conditional CMB visibility of high-redshift X-ray cavities in galaxy clusters}

\author[0009-0002-0373-570X]{Ramananda Santra}
\email[show]{ramananda.santra@icts.res.in}
\correspondingauthor{Ramananda Santra}
\affiliation{International Centre for Theoretical Sciences, Tata Institute of Fundamental Research, Bengaluru, Karnataka 560089, India}

\author[0000-0002-5730-0376]{Samik Mitra}
\email[show]{samik.mitra@icts.res.in}
\affiliation{International Centre for Theoretical Sciences, Tata Institute of Fundamental Research, Bengaluru, Karnataka 560089, India}

\begin{abstract}
At high redshift, the cosmic microwave background (CMB) both cools relativistic leptons and provides seed photons for inverse-Compton (IC) X-ray emission. These two effects link the particle content of a bubble directly to the soft X-ray depression used to identify a cavity. We derive a physical criterion connecting lepton cooling, bubble volume, enthalpy evolution, and 0.5--2 keV X-ray visibility, while separating intrinsic filling of the X-ray depression from detector-dependent cavity recovery. For uncooled electron populations described by a power law with spectral index $s=2.5$, minimum Lorentz factor $100\leq\gamma_{\min}\leq1000$, and maximum Lorentz factor $\gamma_{\max}=10^5$, the measured atmosphere of the testbed high-redshift cluster SPT-CL J2215$-$3537 requires 0.6--1.2\% of the pressure in radiating leptons to fill the soft-band deficit. Across the full tested ranges of $s$, $\gamma_{\min}$, and atmospheric properties, the required fraction is 0.2--3.7\%. Complete loss of this component reduces the enthalpy by less than 3\% relative to an otherwise identical adiabatic bubble at the same pressure. A low-cutoff control with $\gamma_{\min}=10$ and $s=3$ instead requires 18.9\%, identifying the poorly constrained low-energy spectrum as the main physical uncertainty. For a fixed atmosphere and spectrum, the required fraction scales as $(1+z)^{-4}$. The declining-pressure tracks show that the X-ray deficit can cross the adopted detection threshold while the enthalpy deficit remains below 0.4\%; the time of this transition depends on the assumed pressure history. Thus, high-redshift cavity visibility depends on both the surrounding atmosphere and the pressure-carrying particle spectrum. This criterion provides the physical basis for an instrument-specific recovery analysis.
\end{abstract}



\section{Introduction}\label{sec:intro}
Mechanical feedback from accreting supermassive black holes is widely regarded as a key process regulating the thermal evolution of massive galaxies and the hot intracluster medium (ICM) in galaxy clusters \citep{McNamara2007, Fabian2012}. The most direct observational tracers of this feedback are X-ray cavities, where the work done in displacing the surrounding gas provides an estimate of the mechanical energy injected by the central black hole \citep[see][for a review]{larrondo2022}. For a relativistic cavity, this enthalpy is commonly written as $H=4pV$, where $p$ is the bubble pressure, assumed to equal the surrounding ICM pressure, and $V$ is the cavity volume. Combined with a buoyancy, sound-crossing, or refill timescale, this yields an estimate of the time-averaged jet power \citep{Birzan2004, Rafferty2006}. As cavity searches are extended to progressively higher redshifts \citep{larrondo2012, larrondo2015}, the completeness of the detected cavity population becomes increasingly important for reconstructing the cosmic history of radio-mode feedback.

The recent extension of cavity searches toward $z\sim1$ makes the question of completeness increasingly timely. Cavity demographics remain one of the few direct observational routes to the cosmic evolution of radio-mode mechanical power, yet the assumptions inherited from low-redshift studies are least tested in this regime. Radio-faint ghost cavities show that relativistic plasma can outlive its bright radio phase \citep{Birzan2008,Ensslin2002}, while aged lobes and remnants may remain visible through inverse-Compton (IC) X-ray emission \citep{Celotti2004,Mocz2011,HodgesKluck2021}. At high redshift, the stronger IC loss channel also alters radio-lobe energetics \citep{Ghisellini2015}. These studies establish the relevant physical ingredients, but they do not determine how those ingredients modify the X-ray criterion used to identify a cavity.

At high redshift, a basic assumption inherited from low-redshift cavity studies begins to fail: the bubble contents and the X-ray depression used to identify the bubble are no longer independent. Because the CMB energy density, $U_{\rm CMB}$, scales as $(1+z)^4$, the same photon field both increases the IC cooling rate and provides the seed photons for IC X-ray emission. Radiating leptons can therefore lose pressure while replacing part, or even all, of the soft-band thermal signal displaced by the bubble. A cavity may remain energetically important even as its X-ray shadow weakens. For a fixed atmosphere and lepton spectrum, the radiating-lepton fraction required for full soft-X-ray fill-in scales as $(1+z)^{-4}$, so the redshift dependence is explicit. The possibility that IC emission could fill a high-redshift cavity was raised by \citet{Markevitch2009}, while \citet{Hardcastle2026} followed radiative losses in radio-galaxy dynamics. What has remained missing is a direct calculation connecting the lepton pressure, the evolving bubble volume and enthalpy, and the soft X-ray contrast in a measured atmosphere. 

In this paper, we develop an analytical model which follows these quantities with a single accumulated cooling variable, separates physical filling of the soft-band deficit from the observational threshold used to recover a cavity, and determines the required lepton pressure for an observed cluster atmosphere. The reported SPT-CL J2215$-$3537 (hereafter, J2215) feature \citep{Bessa2026} is used only to supply the measured temperature, density, and X-ray emissivity at its projected radius as a testbed; none of the calculations requires the feature itself to be a cavity. Section~\ref{sec:formalism} develops the cooling and contrast model, Section~\ref{sec:results} gives the lepton pressure required to fill the X-ray deficit and illustrates the resulting contrast evolution, and Section~\ref{sec:discussion} discusses the physical limits and observational implications. We adopt the same flat $\Lambda$CDM cosmology as \citet{Bessa2026}, with $H_0=70$ km s$^{-1}$ Mpc$^{-1}$, $\Omega_m=0.3$, and $\Omega_\Lambda=0.7$.

\section{Physical Model}\label{sec:formalism}

\subsection{Model assumptions and cooling evolution}\label{sec:model}

We follow the bubble from detachment, defined as the time at which the jet stops inflating it. The bubble contains two components: radiating leptons (electrons and positrons, hereafter ``electrons''), described by a single homogeneous power-law spectrum, and a non-radiating component whose radiative losses are negligible (for example, relativistic protons). Both are assigned an adiabatic index $\Gamma=4/3$.  Appendix~\ref{app:robustness} separately tests a thermal, non-relativistic non-radiating component with $\Gamma_{\rm nr}=5/3$. The bubble is assumed to remain in pressure balance with its surroundings, and the external pressure along its trajectory is prescribed rather than solved. We neglect particle escape and re-acceleration, and do not model the buoyant trajectory, bubble disruption, or the surrounding pressure profile. Pressure balance requires the bubble to adjust to external pressure changes faster than its contents cool. At the J2215 feature ($z=1.16$, $kT=7.77$ keV; Section~\ref{sec:fillin}), the ambient sound-crossing time across one bubble radius, $R_b\simeq33$ kpc, is about 23 Myr. This is $0.16\,t_{\rm IC}$, where $t_{\rm IC}$ is the inverse-Compton cooling time (Equation~\ref{eq:tic}) of electrons with $\gamma_*\simeq753$, the Lorentz factor near the peak of the 0.5--2 keV IC response (Section~\ref{sec:xray}). This comparison supports approximate pressure balance for the shorter illustrative trajectories, although the approximation becomes less secure for the longest tracks.

Let $p_{\rm det}$ and $V_{\rm det}$ be the bubble pressure and volume at detachment, and define $P(t)\equiv p(t)/p_{\rm det}$ and $\mathcal{V}(t)\equiv V(t)/V_{\rm det}$, so that $P(0)=\mathcal{V}(0)=1$. The adiabatic and radiative aging equation \citep[e.g.,][]{Ensslin2002,Mocz2011} is
\begin{align}
\dot\gamma&=-\frac{\dot{\mathcal V}}{3\mathcal V}\gamma-a(t)\gamma^2,\qquad
a(t)=\frac{4\sigma_{\rm T}}{3m_ec}(U_{\rm CMB}+U_B),\nonumber\\
\gamma(t)&=\frac{A(t)\gamma_0}{1+\Psi(t)\gamma_0},\qquad
A=\mathcal V^{-1/3},\label{eq:characteristic}\\
\Psi(t)&=\int_0^t a(t')\mathcal V(t')^{-1/3}\dd t'.\nonumber
\end{align}
Here $\sigma_{\rm T}$ is the Thomson cross section, $m_e$ is the electron mass, and $c$ is the speed of light. The quantity $\gamma(t)$ is the electron Lorentz factor at time $t$, $\gamma_0$ is its value at detachment, and $a(t)$ is the radiative-loss coefficient. The first term in $\dot{\gamma}$ describes the adiabatic change in electron energy as the bubble expands or contracts, and the second describes inverse-Compton losses against the CMB and synchrotron losses in the magnetic field, whose energy densities are $U_{\rm CMB}$ and $U_B$, respectively. The factor $A(t)$ describes the adiabatic energy change, whereas $\Psi(t)$ measures the accumulated radiative cooling. Equation~(\ref{eq:characteristic}) therefore maps each initial Lorentz factor $\gamma_0$ to its value at time $t$, which we use below to evolve the electron distribution.

At detachment, we assume a power-law electron spectrum,
$$
\frac{\dd N_0}{\dd\gamma_0}=K\gamma_0^{-s},\qquad\gamma_{\min}\leq\gamma_0\leq\gamma_{\max}.
$$
Here $N_0$ denotes the electron number distribution at detachment. The constant $K$ is the normalization, $s$ is the spectral index, and $\gamma_{\min}$ and $\gamma_{\max}$ are the low- and high-energy cutoffs. We define
\begin{align}
I_1&\equiv\int_{\gamma_{\min}}^{\gamma_{\max}}\gamma_0^{1-s}\dd\gamma_0,\nonumber\\
R_1(\Psi)&=\frac{1}{I_1}\int_{\gamma_{\min}}^{\gamma_{\max}}
\frac{\gamma_0^{1-s}}{1+\Psi\gamma_0}\dd\gamma_0.
\label{eq:r1}
\end{align}
The quantity $I_1$ normalises the initial electron energy distribution. With this normalization, $R_1=1$ at detachment, before radiative cooling has acted, and decreases as the electrons lose energy. Thus, $R_1$ measures the fraction of the radiating-electron pressure that remains after radiative cooling, apart from the common adiabatic change associated with the changing bubble volume. We define

$$
f_{e,0}\equiv\frac{p_{e,\rm det}}{p_{\rm det}},
$$

where $p_{e,\rm det}$ is the radiating-electron pressure at detachment, so that $f_{e,0}$ is the fraction of the total bubble pressure carried by these electrons. The remaining fraction, $1-f_{e,0}$, is supplied by the non-radiating component. Because each electron's energy scales as $\gamma$, Equation~(\ref{eq:characteristic}) gives the radiating-electron pressure $p_e=p_{e,\rm det}\mathcal V^{-4/3}R_1(\Psi)$, while the non-radiating pressure is $p_{\rm nr}=(1-f_{e,0})\,p_{\rm det}\mathcal V^{-4/3}$. Pressure balance, $p_e+p_{\rm nr}=p$, then gives
\begin{align}
P&=\mathcal V^{-4/3}D,\qquad D=1-f_{e,0}+f_{e,0}R_1,\nonumber\\
\mathcal V&=\left(\frac{D}{P}\right)^{3/4},\qquad
\dot\Psi=a(t)\left(\frac{P}{D}\right)^{1/4}.
\label{eq:closed}
\end{align}
Here $D$ accounts for the reduction in pressure caused by radiative losses from the electron population. Without radiative cooling, $R_1=1$ and therefore $D=1$, recovering the usual relativistic adiabatic relation $P=\mathcal V^{-4/3}$. As the electrons cool, $R_1<1$ and hence $D<1$, so the bubble volume must adjust to maintain pressure balance. The final equation follows by substituting this volume relation into the definition of the cooling coordinate $\Psi$ in Equation~(\ref{eq:characteristic}).

For any prescribed pressure history $P(t)$, the electron spectrum, bubble volume, and enthalpy thus follow from the single cooling variable $\Psi$. Since $H=4pV$ for contents with $\Gamma=4/3$, the enthalpy relative to its value at detachment, and relative to an otherwise identical adiabatic bubble at the same pressure, is
\begin{equation}
\frac{H}{H_{\rm det}}=P^{1/4}D^{3/4},\qquad
\Crad\equiv\frac{H_{\rm ad}(P)}{H}=D^{-3/4}.
\label{eq:crad}
\end{equation}

Here $H$ is the bubble enthalpy, $H_{\rm det}$ is its value at detachment, and $H_{\rm ad}(P)$ is the enthalpy of an otherwise identical bubble evolved adiabatically to the same external pressure $P$.

The corresponding fractional enthalpy deficit is
\begin{equation}
\Delta_H\equiv1-\frac{H}{H_{\rm ad}}=1-\Crad^{-1}.
\label{eq:deltah}
\end{equation}
At the same ambient pressure, $V/V_{\rm ad}=\Crad^{-1}$. A spherical bubble rising at its terminal speed, where buoyancy balances drag, has $v_b\propto(gR_b)^{1/2}$, where $g$ is the local gravitational acceleration. Since $R_b\propto V^{1/3}$, this gives $v_b/v_{b,\rm ad}=\Crad^{-1/6}$ \citep[e.g.,][]{Kaiser2005}, where $v_b$ and $v_{b,\rm ad}$ are the terminal rise speeds of the radiatively affected and the corresponding adiabatic bubble. A large change in X-ray surface brightness need not imply a large change in the bubble enthalpy, because the radiating leptons carry only a small fraction of the total pressure. For example,
complete loss of a radiating component with $f_{e,0}=0.01$ gives $\Crad=(1-0.01)^{-3/4}=1.008$, or $\Delta_H=0.75$\%, whereas Section~\ref{sec:fillin} shows that a component of this size can fill the soft-X-ray deficit.

We use $\Crad=3$ as a model-validity limit for the illustrative trajectories and exclude crossings occurring beyond this value from the quoted results, because the model does not solve the dynamical response of the bubble to its own radiative losses self-consistently. At $\Crad=3$, the terminal rise speed is already $1-3^{-1/6}\simeq17$\% below its adiabatic value, so the neglected dynamical feedback may no longer be small. Magnetic losses are parameterized by $b_0\equiv U_{B,0}/U_{\rm CMB}$, where $U_{B,0}$ and $U_{\rm CMB}$ are the magnetic and CMB energy densities at detachment. Assuming isotropic flux freezing, $B=B_0\mathcal V^{-2/3}$, where $B_0$ and $B$ are the magnetic-field strengths at detachment and at time $t$, respectively, giving $U_B/U_{\rm CMB}=b_0P/D$. Magnetic pressure is not included in Equation~(\ref{eq:closed}). All main-text evolutionary results adopt $b_0=0$, i.e., IC cooling only.

\subsection{Observed X-ray response}\label{sec:xray}

The cooling solution above determines the evolving lepton spectrum. To connect it to the observable X-ray emission, we calculate the response in a fixed observed band $[E_1,E_2]$, where $E_1$ and $E_2$ are the lower and upper observed-frame photon energies. The full isotropic Thomson response \citep{Blumenthal1970} is
\begin{align}
 W_X(\gamma)=\frac{135}{\pi^4}\int_{y_1}^{\infty}\frac{x^3\dd x}{\e^x-1}
 \Bigg\{&G\!\left[\min\!\left(1,\frac{y_2}{x}\right)\right]
 -G\!\left(\frac{y_1}{x}\right)\Bigg\},
 \label{eq:kernel}
\end{align}
where $x=\epsilon/kT_{\rm CMB}$ is the CMB photon energy $\epsilon$ in units of $kT_{\rm CMB}$, the bracket vanishes for $y_1/x\geq1$ and
\begin{equation}
 y_i=\frac{E_i}{4\gamma^2kT_{\rm CMB,0}},\qquad
 G(\xi)=\frac{\xi^2}{2}+\frac{\xi^3}{9}+\frac{2\xi^3}{3}\ln \xi-\frac{\xi^4}{2}.
 \label{eq:gkernel}
\end{equation}
Here $k$ is Boltzmann's constant, $T_{\rm CMB,0}$ is the present-day CMB temperature, $y_i$ is the dimensionless band edge corresponding to $E_i$, and $\xi$ is the dimensionless scattering variable. The prefactor normalises the all-band response to unity. Since the emitted band edges and $T_{\rm CMB}$ share the same $(1+z)$ factor, the dimensionless response $W_X(\gamma)$ is redshift invariant in a fixed observed band. The IC-to-thermal X-ray ratio ($\RX$, Equation~\ref{eq:contrast}) is not redshift invariant, because it retains $U_{\rm CMB}$ and the evolving thermal denominator. The 0.5--2 keV response peaks at $\gamma\simeq749$, or at $\gamma\simeq762$ after Galactic absorption. We evaluate the characteristic IC cooling time at $\gamma_*\simeq753$, the unabsorbed peak on our numerical grid; the difference is below 2\%. Even for $\gamma_{\max}=10^5$ at $z=1.5$, $4\gamma_{\max}kT_{\rm CMB}/m_ec^2=4.60\times10^{-4}$, so the calculation remains safely in the Thomson regime.

We describe the contribution of each Lorentz factor to the electron pressure through
\begin{equation}
q(\gamma,t)=\frac{1}{p}\frac{\dd p_e}{\dd\ln\gamma},\qquad
\QX(t)=\int W_X(\gamma)\gamma q(\gamma,t)\dd\ln\gamma.
\label{eq:qxdef}
\end{equation}
Here $p_e$ is the radiating-electron pressure, $q(\gamma,t)$ is its contribution per logarithmic Lorentz-factor interval normalised by the total bubble pressure $p$, and $\QX$ is the corresponding X-ray-response-weighted moment. Thus, $\int q\dd\ln\gamma=p_e/p$ rather than unity. At detachment,
\begin{equation}
\QX(0)=f_{e,0}\MX,\qquad
\MX=\frac{\int\gamma^{2-s}W_X(\gamma)\dd\gamma}{I_1}.
\label{eq:mx}
\end{equation}
Here $\MX$ gives the X-ray spectral response per unit radiating-electron pressure at detachment. All numerical moments and figures use the Galactic-screened response. At $\gamma_{\min}=100$, $\MX=101.9$, 104.0, and 67.2 for $s=2.1$, 2.5, and 3.0, respectively. For $s=2.5$, $\MX=184.2$ and 149.0 at $\gamma_{\min}=300$ and 1000. The spectral response $\MX$ depends non-monotonically on $\gamma_{\min}$: raising the cutoff first removes electrons that carry pressure but contribute little soft-band IC emission, so $\MX$ increases, and $\MX$ decreases again once the cutoff reaches the energies that dominate the response. We also define the response-weighted Lorentz factor at detachment
\begin{equation}
\gamma_{\rm eff}\equiv
\frac{\int\gamma^{3-s}W_X(\gamma)\dd\gamma}
{\int\gamma^{2-s}W_X(\gamma)\dd\gamma},
\label{eq:gammaeff}
\end{equation}
which gives the characteristic Lorentz factor of the electrons contributing to the observed X-ray band and generally differs from the kernel maximum $\gamma_*$. Using the cooling relation in Equation~(\ref{eq:characteristic}), the X-ray-weighted moment evolves as
\begin{align}
\QX(t)=\frac{f_{e,0}A}{I_1D}\int_{\gamma_{\min}}^{\gamma_{\max}}
\frac{\gamma_0^{2-s}}{(1+\Psi\gamma_0)^2}
W_X\!\left(\frac{A\gamma_0}{1+\Psi\gamma_0}\right)\dd\gamma_0.
\label{eq:qxevolved}
\end{align}
Because $W_X(\gamma)$ is broad, electrons can cool into the range that contributes strongly to the observed band before later cooling below it.

We next relate this spectral moment to the observed X-ray deficit. Let $\delta_0>0$ be the 0.5--2 keV deficit of an empty cavity relative to the undisturbed atmosphere. If $I_{\rm disp}$ is the X-ray intensity displaced by the cavity and $I_{\rm IC}$ is the IC intensity produced inside the cavity,
\begin{align}
\delta_X&=\delta_0(1-\RX),\qquad \RX\equiv I_{\rm IC}/I_{\rm disp},\\
\RX&=\FIC\frac{4\sigma_{\rm T}U_{\rm CMB}p}{m_ecn_en_{\rm H}\Lambda_X}\QX.
\label{eq:contrast}
\end{align}
Here $\delta_X$ is the remaining X-ray deficit, $\FIC$ is the geometric correction for the IC-emitting path length, $n_e$ and $n_{\rm H}$ are the electron and hydrogen number densities of the surrounding gas, and $\Lambda_X$ is its 0.5--2 keV emissivity coefficient. When the IC and displaced thermal emission arise from the same projected region, the common $(1+z)^{-4}$ surface-brightness dimming cancels in their ratio. The remaining redshift dependence enters through $U_{\rm CMB}$ and the atmosphere. The condition $\RX=1$ means that the IC emission exactly replaces the displaced thermal emission in the 0.5--2 keV energy flux. It need not give equal detector counts, because the instrument response weights the IC and thermal spectra differently.

The foreground-absorbed 0.5--2 keV emissivity per $n_en_{\rm H}$ is
\begin{equation}
\Lambda_X=\int_{(1+z)E_1}^{(1+z)E_2}
\lambda_E^{\rm APEC}(T,Z)\e^{-N_{\rm H}\sigma[E/(1+z)]}\dd E,
\label{eq:lambda}
\end{equation}
where $\lambda_E^{\rm APEC}(T,Z)$ is the APEC spectral emissivity at gas temperature $T$ and metallicity $Z$, $N_{\rm H}$ is the Galactic hydrogen column density, and $\sigma(E)$ is the photoelectric absorption cross section per hydrogen atom. We use AtomDB/APEC 3.1.3, $Z=0.3Z_\odot$ on the abundance scale of \citet{Anders1989}, and $n_{\rm H}=0.83n_e$ \citep{Smith2001a,Foster2012}. The adopted absorbed emissivities, computed with pyatomdb 1.2.0/CIESession for $z=1.16$ and $N_{\rm H}=1.01\times10^{20}\ {\rm cm^{-2}}$, are $\Lambda_X=(7.83,8.02,8.15)\times10^{-24}\ \mathrm{erg\,cm^3\,s^{-1}}$ at $kT=(6.33,7.77,10.29)$ keV. For consistency, all quoted moments, thresholds, and figures include the same Galactic absorption in the IC emission. Relative to the unabsorbed response, a neutral-gas transmission matched to the adopted $N_{\rm H}$ and \cite{Anders1989} abundance scale lowers $\MX$ by 2.2--3.0\% over the displayed $\gamma_{\min}$ range without changing the qualitative behaviour. Because the absorption depends on observed energy, the screened fixed-band response remains redshift invariant. We retain the abundance scale used in the source analysis.

The factor $\FIC$ corrects the IC-to-thermal ratio for the thermal emission displaced along the cavity line of sight. For a uniformly emitting projected fraction $\phi$ of a spherical cavity of radius $R_b$, centered at projected radius $b$ and line-of-sight coordinate $\ell_0$, in an isothermal $\beta$-model, the factor can be written as:
\begin{equation}
\FIC=
\phi
\left[
\frac{1}{2R_b}
\int_{\ell_0-R_b}^{\ell_0+R_b}
\left(
\frac{1+(b^2+\ell^2)/r_c^2}
{1+(b^2+\ell_0^2)/r_c^2}
\right)^{-3\beta}
\dd \ell
\right]^{-1}.
\label{eq:ficbeta}
\end{equation}
Here $\phi$ is the projected IC-emitting fraction of the cavity, $r_c$ is the $\beta$-model core radius, and $\beta$ is its slope parameter. The term in brackets is the mean thermal emissivity along the cavity chord relative to its value at the cavity centre. Direct integration for our testbed cluster SPT-CL J2215$-$3537, using $R_b\simeq33$ kpc, $b\simeq230$ kpc, and its published double-$\beta$ density profile, gives $0.9817\leq\FIC/\phi\leq1.0146$ over $|\ell_0|\leq b$. Thus, for this system the geometric correction is only at the percent level.

The radiating-lepton pressure fraction used here is local: it refers to the plasma that produces the IC emission, and we do not infer a volume-averaged value. In the time-dependent examples, we assume the emitting plasma fills the whole cavity, so $\phi=1$; a smaller covering fraction is used only in the instantaneous fill-in calculation. We exclude any compressed rim from $I_{\rm disp}$, since its emission is instead accounted for separately, in the two-dimensional surface-brightness model used to detect the cavity.

The soft-band thermal deficit is exactly filled when $\RX=1$. This condition is evaluated at a single instant and does not require the evolutionary model of Section~\ref{sec:model} or a prescribed pressure history. For an idealised detection threshold $S_{\min}\equiv\delta_{\rm det}/\delta_0$, where $\delta_{\rm det}$ is the minimum deficit required for detection, the cavity satisfies the detection criterion when
\begin{equation}
\RX<R_{\rm crit}\equiv1-S_{\min},
\label{eq:selection}
\end{equation}
and fails it when $\RX>R_{\rm crit}$. We refer to a crossing from $\RX>R_{\rm crit}$ to $\RX<R_{\rm crit}$ as a downward, or recovery, crossing. The condition $\RX=1$ describes complete filling of the energy deficit in the chosen band, whereas Equation~(\ref{eq:selection}) is only a simplified detection threshold, not a calibrated survey completeness function. For reference, Appendix Figure~\ref{fig:physical}(b) also shows $\mathcal R_{X,b}=3$ as an illustrative IC-bright state, not as a boundary.

\begin{figure*}
\centering
\includegraphics[width=\textwidth]{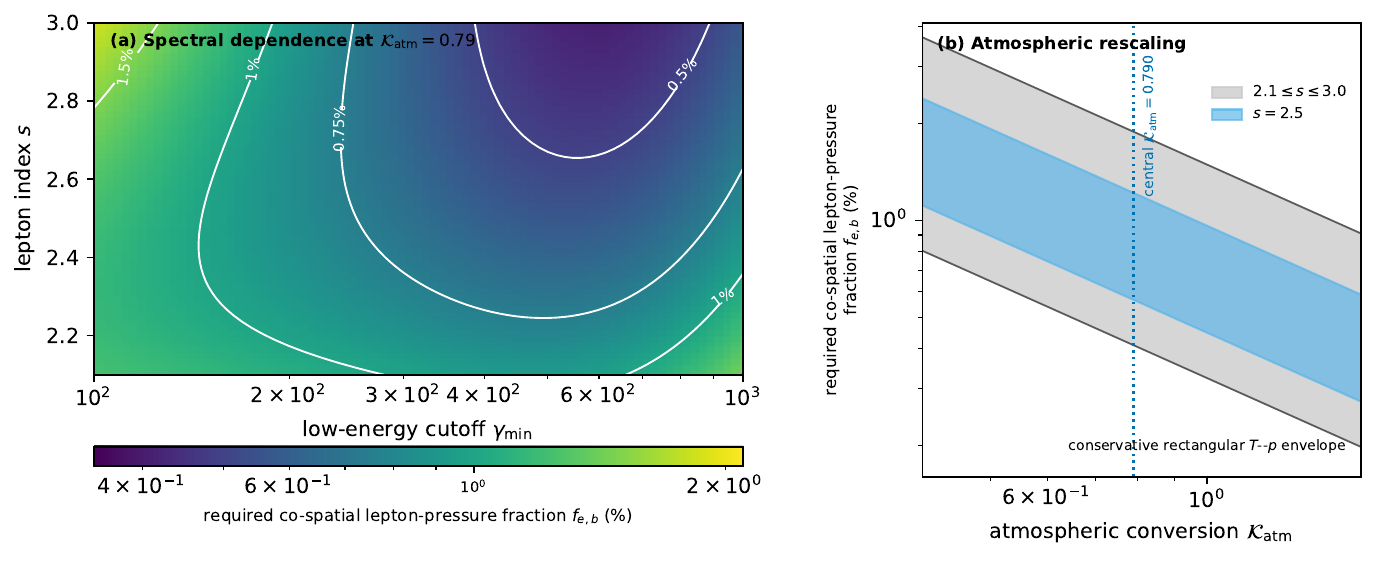}
\caption{Dependence of the soft-X-ray fill-in threshold on the electron spectrum and the cluster atmosphere. \textit{Left:} The radiating-lepton pressure fraction
required for $\mathcal R_{X,b}=1$ in the central J2215 atmosphere, with $\Katm = 0.790$ and $\FIC=1$. The contours give the required fraction in per cent as a function of the spectral index $s$ and low-energy cutoff
$\gamma_{\min}$. \textit{Right:} The required pressure fraction as the atmospheric factor $\Katm$ varies over the conservative range 0.403--1.635. The blue band shows $s=2.5$, while the grey band includes
$2.1\leq s\leq3.0$; both use $100\leq\gamma_{\min}\leq1000$ and $\gamma_{\max}=10^5$. The vertical line marks the central J2215 value.}
\label{fig:robustness}
\end{figure*}

\section{Results}\label{sec:results}

For the instantaneous fill-in calculation, the benchmark state is a local region containing an uncooled power-law electron spectrum. In the time-dependent examples, the same state defines the start of the trajectory, so $f_{e,b}=f_{e,0}$ and $P(0)=1$. Here $f_{e,b}$ is the local radiating-lepton pressure fraction at the benchmark state, whereas $f_{e,0}$ denotes the same fraction at the start of an evolutionary trajectory. The IC-to-thermal X-ray ratio at the benchmark state is
\begin{equation}
\mathcal R_{X,b}=\FIC f_{e,b}\MX\Katm,\qquad
\Katm\equiv\frac{4\sigma_{\rm T}U_{\rm CMB}p}
{m_ecn_en_{\rm H}\Lambda_X}.
\label{eq:katm}
\end{equation}
Here $\Katm$ contains the dependence on the local gas properties and the CMB energy density; the other three factors are the geometric factor $\FIC$, the radiating-lepton pressure fraction $f_{e,b}$, and the spectral response $\MX$ (Equation~\ref{eq:mx}). The main instantaneous fill-in results are summarized in Table~\ref{tab:results}.

\begin{table}[t]
\caption{Soft-X-ray fill-in thresholds. The second column gives the required product $\FIC f_{e,b}$ for $\RX=1$. The final column gives $\Crad$ after complete loss of the corresponding co-spatial lepton component, assuming $\FIC=1$.}
\label{tab:results}
\centering
\scriptsize
\setlength{\tabcolsep}{2.5pt}
\begin{tabular}{@{}lcc@{}}
\hline
Case & Required $\FIC f_{e,b}$ & $\Crad$ \\
\hline
Central atmosphere, $s=2.5$ & 0.57--1.22\% & 1.004--1.009 \\
Atmospheric $T$--$p$ range, $s=2.5$ & 0.27--2.39\% & 1.002--1.018 \\
Full spectral and $T$--$p$ range & 0.20--3.70\% & 1.001--1.029 \\
Low-cutoff control & 18.9\% & 1.170 \\
\hline
\end{tabular}
\end{table}

\subsection{Pressure required for X-ray fill-in}\label{sec:fillin}

We first evaluate the fill-in condition using the measured atmosphere of SPT-CL J2215$-$3537 at $z=1.16$, which we use as the testbed for the model. At the projected radius of the feature, the azimuthally averaged radial bin gives $kT=7.77^{+2.52}_{-1.44}$ keV and $p=0.06^{+0.02}_{-0.01}\mathrm{keV\,cm^{-3}}$ \citep{Bessa2026}. The pressure is derived from the deprojected density profile using $p=1.92n_ekT$, giving $n_e=4.02\times10^{-3}\ {\rm cm^{-3}}$ at the central values and $n_{\rm H}=0.83n_e$. Using the central temperature, pressure, and X-ray emissivity, we obtain $\Katm=0.790$, which scales as
\begin{align}
\frac{\Katm}{0.790}
&=\left(\frac{T}{7.77\ {\rm keV}}\right)^2
\left(\frac{0.06\ \mathrm{keV\,cm^{-3}}}{p}\right)\nonumber\\
&\quad\times\left(\frac{8.02\times10^{-24}\mathrm{erg\,cm^3\,s^{-1}}}{\Lambda_X}\right).
\label{eq:katmrange}
\end{align}
Combining the quoted upper and lower limits on temperature and pressure with the corresponding emissivities gives the conservative range
\begin{equation}
\Katm=0.403\text{--}1.635.
\label{eq:katmenvelope}
\end{equation}
Because the covariance between temperature and spectral normalization is unavailable, this range is not a statistical confidence interval.

For $s=2.5$, $100\leq\gamma_{\min}\leq1000$, and
$\gamma_{\max}=10^5$, the central atmosphere requires
\begin{equation}
 \FIC f_{e,b}=(0.57\text{--}1.22)\times10^{-2}
 \qquad (\Katm=0.790)
 \label{eq:access1}
\end{equation}
to fill the thermal deficit in the soft X-ray band. Figure~\ref{fig:robustness}(a) shows how this threshold depends on the electron spectrum. For fixed $s$, the required pressure fraction first decreases as $\gamma_{\min}$ is increased, reaches a minimum at intermediate $\gamma_{\min}$, and then rises again.
This behaviour produces a broad minimum in the required pressure
fraction rather than a monotonic dependence on $\gamma_{\min}$. Across $2.1\leq s\leq3.0$, the central-atmosphere threshold spans 0.41--1.88\% over the adopted spectral range. The threshold also depends on the surrounding atmosphere, as shown in Figure~\ref{fig:robustness}(b). Increasing $\Katm$ increases the IC emission relative to the displaced thermal emission and therefore lowers the required lepton-pressure fraction. The blue band shows the $s=2.5$ cases, while the wider grey band
includes the full range $2.1\leq s\leq3.0$; the vertical dotted line marks the central value $\Katm=0.790$. Thus, both the electron spectrum and the local ICM properties change the numerical threshold, but the required radiating
pressure remains at the percent level around the measured J2215 atmosphere.

Including the atmospheric limits broadens the $s=2.5$ result to 0.27--2.39\%. Over the full adopted spectral domain, $2.1\leq s\leq3.0$ and $100\leq\gamma_{\min}\leq1000$, the threshold becomes
0.20--3.70\% across the conservative temperature--pressure
range (Figure~\ref{fig:robustness}, right). Even at the upper end, complete loss of this co-spatial lepton component reduces the enthalpy by only 2.8\% relative to a same-pressure adiabatic bubble. At fixed $T$, $p$, $\Lambda_X$, spectrum, and overlap, the redshift dependence follows directly from $U_{\rm CMB}$. Writing $f_{e,\rm fill}$ for the radiating-lepton pressure fraction required for full soft-X-ray fill-in, we have:
\begin{equation}
 f_{e,\rm fill}(z)=f_{e,\rm fill}(1.16)
 \left(\frac{2.16}{1+z}\right)^4.
 \label{eq:redshiftscaling}
\end{equation}
For the central $s=2.5$ range, a 1\% component reaches full soft-X-ray fill-in at $z\simeq0.87$--1.27. Across the full adopted spectral and atmospheric range, the corresponding interval is $z\simeq0.44$--2.00. 
These values show only the CMB redshift scaling at fixed atmospheric properties; they are not an evolutionary model for the cluster atmosphere.

The low-energy spectrum is the main remaining 
spectral uncertainty. A control spectrum with
$\gamma_{\min}=10$ and $s=3$ requires 18.9\% in the central atmosphere
because most of its pressure lies below the soft-X-ray response. Such
populations are also more sensitive to Coulomb losses and to departures
from an ultrarelativistic equation of state during cooling. We therefore
adopt $\gamma_{\min}\geq100$ as the primary analysis range, while retaining $\gamma_{\min}=10$ as a
low-energy control case. Partial projected coverage raises the required
local fraction approximately as $\phi^{-1}$; a global pressure fraction
cannot be inferred without a multiphase volume model.

\begin{figure*}
\centering
\includegraphics[width=\textwidth]{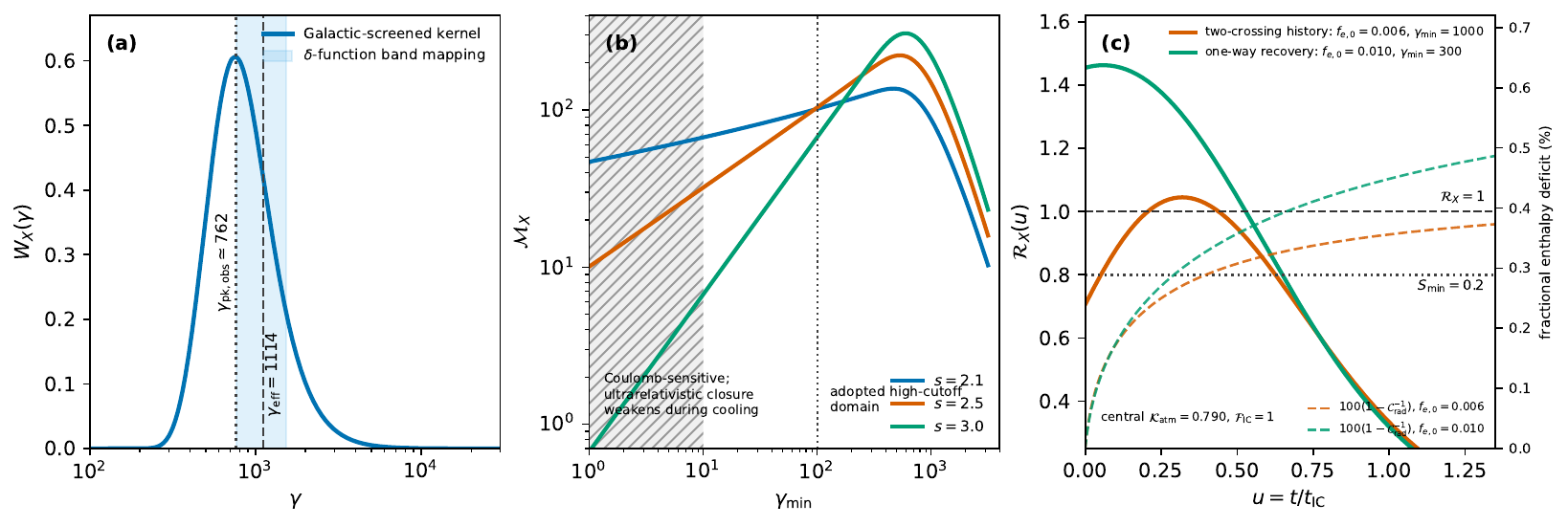}
\caption{Soft-X-ray response and evolution of the cavity signal. \textit{Left:} Galactic-absorbed Thomson response $W_X(\gamma)$ for IC/CMB emission
in the observed 0.5--2 keV band. The response peaks near
$\gamma\simeq762$, while $\gamma_{\rm eff}=1114$ for
$(s,\gamma_{\min})=(2.5,100)$. The shaded interval shows the electron energies selected by the delta-function approximation. \textit{Middle:} The spectral response per unit radiating-lepton pressure, $\MX$, as a function of $\gamma_{\min}$ for $s=2.1$, 2.5, and 3.0. Removing low-energy electrons initially increases the X-ray emission per unit pressure, but the response falls once the cutoff reaches the electrons that contribute strongly to the observed band. \textit{Right:} Two example trajectories in the central J2215 atmosphere. One crosses the adopted detection threshold twice, whereas the other crosses it once as the electrons cool. The right axis gives the fractional enthalpy deficit, which remains below 0.4\% at the downward crossings. Both trajectories use $s=2.5$, $\gamma_{\max}=10^5$,
$P=(1+u)^{-1}$, $b_0=\beta_X=0$, $\FIC=1$, and $S_{\min}=0.2$.}
\label{fig:response}
\end{figure*}

\subsection{Evolution of the X-ray surface brightness}

The instantaneous fill-in threshold above does not require a bubble trajectory. To illustrate how the X-ray emission may evolve after detachment, we use
\begin{align}
 t_{\rm IC}(\gamma_*,z)&\equiv\frac{3m_ec}{4\sigma_{\rm T}U_{\rm CMB}\gamma_*}\nonumber\\
 &=194\ {\rm Myr}
 \left(\frac{\gamma_*}{753}\right)^{-1}
 \left(\frac{1+z}{2}\right)^{-4},\label{eq:tic}\\
 P(u)&=\left(1+\frac{u}{\Theta_p}\right)^{-\eta},
 \qquad u=t/t_{\rm IC},\label{eq:pressurepath}\\
 \frac{T^2/\Lambda_X}{T_0^2/\Lambda_{X,0}}&=P^{\beta_X}.
 \label{eq:betax}
\end{align}
Here $u=t/t_{\rm IC}$ is the dimensionless time, while $T_0$ and $\Lambda_{X,0}$ are the temperature and X-ray emissivity at detachment. The parameter $\Theta_p=\tau_p/t_{\rm IC}$, where $\tau_p$ is the pressure-evolution timescale, sets the timescale over which the pressure changes; $\eta$ controls the decline rate, and $\beta_X=\dd\ln(T^2/\Lambda_X)/\dd\ln P$ describes the thermal response. The ranges $\Theta_p=0.5$--2, $\eta=0.5$--1.5, and $\beta_X=-0.3$ to 0.3 sample declining pressure histories; they are not fitted to J2215. For fixed projected overlap,
\begin{equation}
 \RX(u)=\mathcal R_{X,0}
 \frac{\QX(u)}{f_{e,0}\MX}
 P(u)^{\beta_X-1}.
 \label{eq:rxevolution}
\end{equation}
Here $\mathcal R_{X,0}$ is the IC-to-thermal X-ray ratio at
detachment. If the overlap changes, the right-hand side is multiplied by
$\FIC(u)/\FIC(0)$.

The broad Thomson response reveals leptons over a wider energy range than the
delta-function estimate (Figure~\ref{fig:response}, left). As the population
cools, electrons initially above the energies that contribute most strongly
to the 0.5--2 keV band can move into this range before later cooling below it. Consequently, the soft-X-ray IC emission can increase during the early
stages of cooling even though the electron population is continuously losing
energy. At later times, the electrons that contribute most strongly to the
observed band cool to lower energies and the IC emission decreases. This
allows a non-monotonic X-ray evolution without requiring a comparable change
in the total bubble enthalpy.

The two example trajectories show how this spectral evolution produces
different X-ray histories in the measured central atmosphere of J2215 (Figure~\ref{fig:response}, right).
Both adopt $s=2.5$, $\gamma_{\max}=10^5$, $P=(1+u)^{-1}$,
$b_0=\beta_X=0$, and $\FIC=1$. With $f_{e,0}=0.006$ and
$\gamma_{\min}=1000$, the IC-to-thermal ratio begins at
$\mathcal R_{X,0}=0.706$, rises as electrons cool into the
soft-X-ray-producing energy range, and then decreases again as they cool
below it. It therefore crosses the adopted threshold twice, with the downward crossing at $u_{\rm re}=0.626$, where $u_{\rm re}$ denotes the dimensionless time of the downward recovery crossing. With $f_{e,0}=0.010$ and
$\gamma_{\min}=300$, the ratio begins at $\mathcal R_{X,0}=1.455$ and
decreases after detachment, crossing the same threshold only once at
$u_{\rm re}=0.648$. In this case, a larger fraction of the initial
electron population already contributes efficiently to the observed band, so
there is no comparable initial rise. We refer to the first case as a two-crossing history and to the second as a one-way history. The enthalpy ratios at these downward
crossings are 1.0033 and 1.0039, corresponding to deficits below 0.4\%.
Thus, the X-ray visibility can change substantially while the
mechanical energy of the bubble remains nearly unchanged.
 
The wider parameter grid also includes $S_{\min}=0$ and 0.5. The following interval refers specifically to the selected grid subset, rather than to the two illustrative trajectories quoted above. Restricting this grid subset to $b_0=0$, $\gamma_{\min}\geq100$, cases with a finite downward crossing, and $\Crad\leq3$ gives $u_{\rm re}=0.70$--1.82, corresponding to roughly
0.06--0.35 Gyr over $z=1$--1.5. These are crossing times within the model and depend partly on the prescribed pressure history; they are not measured cavity ages. A two-crossing history additionally requires the cavity to remain coherent until the downward recovery crossing. Defining $t_{\rm re}\equiv u_{\rm re}t_{\rm IC}$, this requires
\begin{equation}
 t_{\rm re}<t_{\rm dis},
 \label{eq:survival}
\end{equation}
where $t_{\rm dis}$ is the disruption or loss-of-coherence time. We impose this condition but do not calculate $t_{\rm dis}$. We evaluate $U_{\rm CMB}$ at the observed redshift throughout; allowing it to evolve would change the numerical crossing times but not their ordering. Backward evolution over 0.35 Gyr at $z=1.16$ raises $U_{\rm CMB}$ by 22\% in the adopted cosmology, so the longest times should not be read as precise ages. The full parameter grid and its numerical controls are described in Appendix~\ref{app:robustness}.

\section{Discussion and Conclusions}\label{sec:discussion}

The result separates two quantities that cavity analyses often treat together. A percent-level radiating-lepton component can fill the X-ray
deficit in the soft X-ray band even though it contributes only a small fraction of the total bubble pressure. The bubble may therefore remain mechanically important even when its X-ray depression becomes weak or disappears. If the X-ray surface-brightness depression is not identified, the cavity cannot enter a standard X-ray cavity analysis, and the corresponding $4pV$ estimate is never made.

The atmospheric factor $\Katm$ and spectral moment $\MX$ determine how much radiating-lepton pressure is required for X-ray fill-in. $\Katm$ increases with $U_{\rm CMB}$ and, at fixed observed-band emissivity, approximately as $kT/n_e$, so the required lepton fraction decreases rapidly with redshift and also decreases in hotter or lower-density gas. $\MX$ measures how efficiently the electron population produces IC emission in the observed band per unit lepton pressure. The quoted percent-level range therefore applies to the tested uncooled power laws with $100\leq\gamma_{\min}\leq1000$ and $\gamma_{\max}=10^5$. Spectral curvature, spatial gradients, and mixed plasma ages will broaden this behaviour, whereas a low-cutoff spectrum can carry substantial pressure while remaining inefficient in soft-band IC emission.

The time-dependent examples show that the X-ray visibility can change while the bubble enthalpy remains nearly unchanged. A two-crossing history additionally requires $t_{\rm re}<t_{\rm dis}$ and little reacceleration over the same interval. Whether such a history occurs in a physical bubble will therefore depend on disruption, entrainment, mixing, magnetic draping, and turbulence \citep{Kaiser2005,Bourne2023}. The prescribed pressure paths are not source fits, and tracks reaching $u\gtrsim3$ are less reliable because pressure balance and the neglect of dynamical feedback (Section~\ref{sec:model}) become less accurate at late times. Excluding these late stages changes neither the instantaneous fill-in threshold nor the early one-way behaviour.

Coulomb and magnetic losses set additional limits on the time-dependent examples. At the lowest Lorentz factor reached in the selected tests, $\gamma\simeq60$, Coulomb and IC losses are equal for an internal thermal-plasma number density $n_{\rm int}=7.7\times10^{-4}\ {\rm cm^{-3}}$, and the Coulomb-loss time satisfies $t_{\rm Coul}\geq3t_{\rm IC}$ for $n_{\rm int}\lesssim2.5\times10^{-4}\ {\rm cm^{-3}}$ \citep{Sarazin1999}. The evolutionary tracks are therefore most applicable to cavities with low internal gas density. At $z=1.16$, the equivalent CMB magnetic field is $B_{\rm CMB}=15.1\ \mu{\rm G}$. Fields of 1--5 $\mu$G give $U_B/U_{\rm CMB}=0.004$--0.11, so IC losses remain dominant and these fields can be treated as tests of additional synchrotron cooling. At $10\ \mu$G, $U_B/U_{\rm CMB}=0.44$, so synchrotron losses are no longer negligible, although $U_B/p\simeq0.04$ at the benchmark pressure. A fully self-consistent calculation at such field strengths would also need to include magnetic pressure in Equation~\eqref{eq:closed}. For $\gamma\sim10^3$, the corresponding observed synchrotron frequency is about 1.7--8.6 MHz for $B=1$--5 $\mu$G, below the terrestrial ionospheric cutoff. Thus, the absence of  radio emission would not by itself rule out the electron population considered here.

The calculation establishes when IC/CMB emission can fill the 0.5--2 keV X-ray deficit before applying the detector response. Translating that threshold into a prediction for observed high-redshift cavity samples requires a source model together with the instrument response, orientation, partial filling, spatially varying spectra, point spread function (PSF), background, and exposure effects. Establishing a population bias further requires the distribution of bubble properties. The present result provides the physical input for those analyses: at fixed atmosphere, geometry, and redshift, stronger IC/CMB emission should be associated with a weaker X-ray cavity deficit. Failure to find such a relation would constrain the prevalence of this channel under the assumed spectra. More broadly, the result connects radio-faint ghost plasma \citep{Birzan2008,Ensslin2002} and IC/CMB-bright aged lobes \citep{Celotti2004,Mocz2011,HodgesKluck2021} to the observable used in X-ray cavity calorimetry. High-redshift cavity visibility therefore depends not only on geometry, but also on the surrounding atmosphere and the particle content of the bubble.

\begin{acknowledgments}
We thank the anonymous referee for their constructive comments that have improved the overall clarity and presentation of the paper. R.S. and S.M. acknowledge support from the Department of Atomic Energy, Government of India, under project no. RTI4019. Generative-AI tools were used for language editing and code debugging; the authors verified the scientific content, calculations, references, and final text.
\end{acknowledgments}

\software{astropy \citep{2013A&A...558A..33A,2018AJ....156..123A,2022ApJ...935..167A}, NumPy \citep{Harris2020}, SciPy \citep{Virtanen2020}, Matplotlib \citep{Hunter2007}, pyatomdb 1.2.0, xraydb 4.5.8}


\appendix

\section{Model Tests and Numerical Checks}\label{app:robustness}

\subsection{Crossing behaviour and projected filling}

\begin{figure*}[t]
\centering
\includegraphics[width=\textwidth]{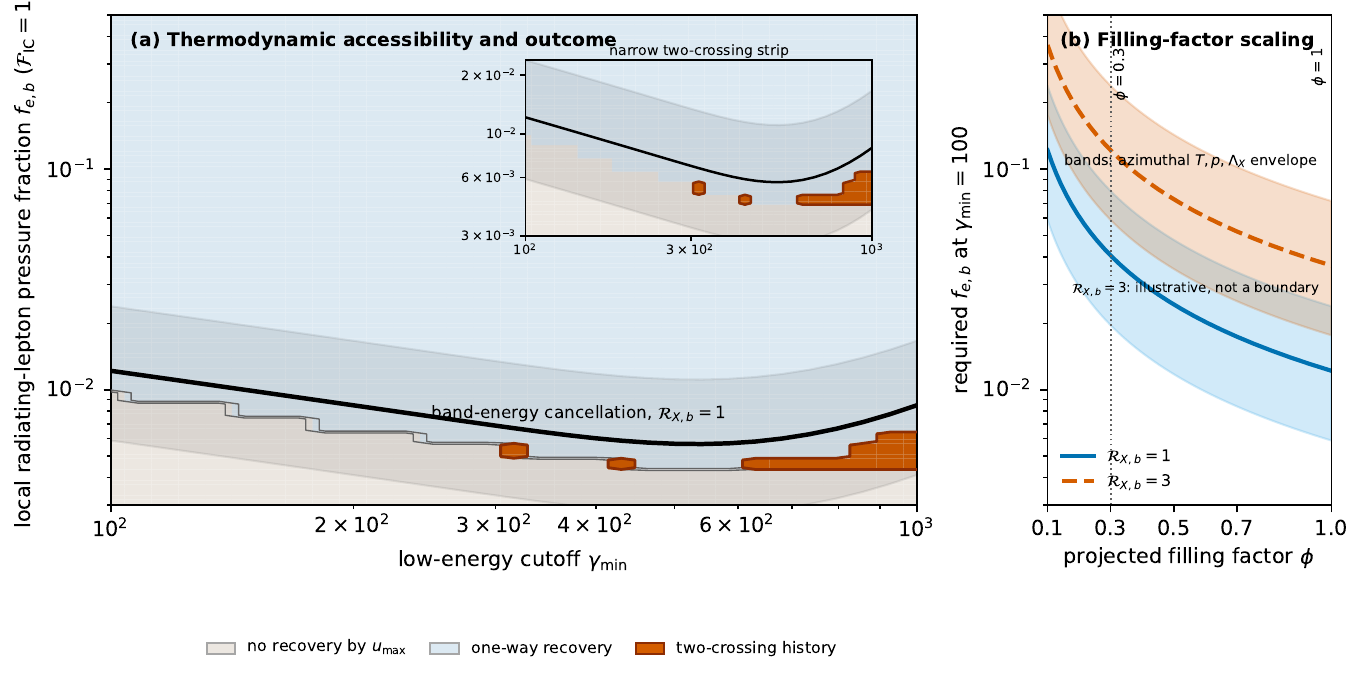}
\caption{Tests of the X-ray evolution and projected filling. \textit{Left:} The crossing behaviour as a function of the low-energy cutoff
$\gamma_{\min}$ and the local radiating-lepton pressure fraction for
$s=2.5$, $S_{\min}=0.2$, $\gamma_{\max}=10^5$, and
$(\Theta_p,\eta,b_0,\beta_X)=(1,1,0,0)$. The orange region contains histories that cross the detection threshold twice, while the blue region
contains cases with one downward crossing. The black curve marks full
soft-X-ray fill-in, $\mathcal R_{X,b}=1$, and the grey band shows the
measured atmospheric range. Cases for which $\Crad>3$ at the downward recovery crossing lie beyond the adopted model-validity limit and are excluded from the quoted grid-based crossing-time interval. \textit{Right:} Required local radiating-lepton pressure fraction at $\gamma_{\min}=100$ as the projected emitting fraction $\phi$ is varied. The blue band gives the full-fill-in condition. The $\mathcal R_{X,b}=3$ band is shown only as an illustrative IC-bright case.}
\label{fig:physical}
\end{figure*}

The instantaneous fill-in thresholds do not depend on a prescribed bubble trajectory. The pressure-history grid is used only to test how the X-ray evolution changes when the model parameters are varied. Figure~\ref{fig:physical}(a) classifies each point of this grid by the resulting crossing behaviour. One-way crossings, in which the X-ray deficit becomes detectable once and stays detectable, occupy a broad part of the parameter space. Two-crossing histories, in which the deficit is first filled and then reappears, are confined to the narrow strip shown in the inset, and therefore require a more specific combination of spectrum and pressure history. A third class of tracks shows no recovery crossing by $u_{\max}=6$, where $u_{\max}$ is the maximum dimensionless integration time. The black curve marks the instantaneous fill-in condition $\mathcal R_{X,b}=1$ from Section~\ref{sec:results}, so the map shows where the time-dependent behaviour sits relative to that threshold. Figure~\ref{fig:physical}(b) shows the separate effect of partial projected coverage, which increases the required local lepton-pressure fraction approximately as $\phi^{-1}$.

\subsection{Equation of state and age mixing}

For radiating leptons with adiabatic index $\Gamma_e=4/3$ and non-radiating support with adiabatic index $\Gamma_{\rm nr}$,
\begin{equation}
 P=(1-f_{e,0})\mathcal V^{-\Gamma_{\rm nr}}
 +f_{e,0}R_1(\Psi)\mathcal V^{-4/3}.
 \label{eq:eosclosure}
\end{equation}
For each $P$ and $\Psi$, this equation gives the bubble volume
$\mathcal V$. The total enthalpy is
\begin{equation}
 H=\left(4p_e+\frac{\Gamma_{\rm nr}}{\Gamma_{\rm nr}-1}p_{\rm nr}\right)V.
 \label{eq:eosenthalpy}
\end{equation}
Here $p_e$ and $p_{\rm nr}$ are the radiating-lepton and
non-radiating pressures, respectively. Taking $\Gamma_{\rm nr}=5/3$ represents thermal, non-relativistic support. The two-crossing behaviour is still present in this case and extends to somewhat lower $f_{e,0}$. The electron-dominated limit is unchanged because the non-radiating pressure vanishes as $f_{e,0}\to1$.

We also test whether a spread in plasma ages changes the X-ray
evolution. We average the response as
\begin{equation}
 \overline{\QX}(u)=\frac{\int\QX(ru)w(r)\dd r}{\int w(r)\dd r}.
 \label{eq:agemix}
\end{equation}
Here $r$ rescales the effective plasma age and $w(r)$ is the corresponding weight. For a top-hat distribution, let $\Delta_r$ denote the fractional half-width, so that $r\in[1-\Delta_r,1+\Delta_r]$ with uniform weight; for a lognormal distribution, $\Delta_r$ denotes the coefficient of variation at unit mean. For $\Delta_r=0.5$, both choices preserve the reference crossings. This test includes only the effect of mixing different plasma ages on the X-ray response; it does not model their dynamics separately.

\subsection{Parameter range and numerical checks}

The parameter grid varies $s=2.1$, 2.5, 3.0;
$\gamma_{\min}=10$, 100, 300, 1000; $\Theta_p=0.5$, 2;
$\eta=0.5$, 1.5; $b_0=0$, 1; and $\beta_X=-0.3$, 0.3, together with the
reference pressure path. The quoted grid-based crossing interval uses only $b_0=0$, $\gamma_{\min}\geq100$, cases with a finite downward crossing, and $\Crad\leq3$. The two illustrative trajectories in Figure~\ref{fig:response} are reported separately and do not define this interval. The main parameter map fixes $s=2.5$, $\FIC=1$,
$S_{\min}=0.2$, and $u_{\max}=6$. We also repeat the calculation with
$S_{\min}=0$ and 0.5, different pressure paths, $u_{\max}=10$, and half the
integration step. These runs test whether the results depend strongly on the
chosen numerical setup. Cases with $b_0=1$ are not included in the
quoted crossing-time range because Equation~\eqref{eq:closed} does not include
magnetic pressure. For $b_0=1$, $U_B/p\simeq0.094$ initially and reaches
0.41 by $\Crad=3$. The adopted $\gamma_{\max}=10^5$ has the largest effect when $s$ is close to 2. Varying $\gamma_{\max}$ from $10^4$ to $10^6$ changes the central fill-in threshold by at most 44\% over the tested grid,
with the largest change at $s=2.1$ and $\gamma_{\min}=1000$; the
largest changes are 24\% for $s=2.5$ and 9\% for $s=3$.
The main-text range therefore applies to the adopted value
$\gamma_{\max}=10^5$ and does not include uncertainty in this upper cutoff.

Several numerical checks were also carried out. The kernel satisfies
the analytic all-band normalization to $4\times10^{-16}$ and agrees with
direct quadrature to a maximum relative interpolation error
$2.23\times10^{-3}$. Two independent methods for evolving $\Psi$
agree to $3.7\times10^{-7}$. Extending $u_{\max}$ from 6 to 10
does not change the type of crossing history, and halving the step
shifts the reference $u_{\rm re}$ by less than $10^{-3}$. Direct integration
of Equation~\eqref{eq:ficbeta} gives
$0.9817\leq\FIC/\phi\leq1.0146$ across the two J2215 $\beta$ components.
The screened-kernel calculation agrees with the direct moment
calculation, and including Galactic absorption does not change the crossing
behaviour.


\bibliography{sample701}{}
\bibliographystyle{aasjournalv7}



\end{document}